\documentclass{article}
\usepackage{graphicx}
\usepackage{subcaption}
\newlength{\twosubht}
\newsavebox{\twosubbox}
\usepackage{numprint}
\usepackage{subcaption}
\usepackage{caption}
\usepackage{float}
\usepackage{amsmath}
\usepackage{authblk}
\usepackage[round]{natbib}   

\providecommand{\keywords}[1]
{
  \small	
  \textbf{\textit{Keywords---}} #1
}
\title{Unravelling the Spatial Properties of Individual Mobility Patterns using Longitudinal Travel Data}

\date{}

\author[1,2]{Oded Cats}
\author[2]{Francesco Ferranti}
\affil[1]{Email: o.cats@tudelft.nl, Department of Transport and Planning, Delft University of Technology, Delft, The Netherlands;}
\affil[2]{Division of Transport Planning, KTH Royal Institute of Technology, Stockholm, Sweden}

\begin{document}

\maketitle

\begin{abstract}
The analysis of longitudinal travel data enables investigating how mobility patterns vary across the population and identify the spatial properties thereof. The objective of this study is to identify the extent to which users explore different parts of the network as well as identify distinctive user groups in terms of the spatial extent of their mobility patterns. To this end, we propose two means for representing spatial mobility profiles and clustering travellers accordingly. We represent users patterns in terms of zonal visiting frequency profiles and grid-cells spatial extent heatmaps. We apply the proposed analysis to a large-scale multi-modal mobility data set from the public transport system in Stockholm, Sweden. We unravel three clusters - locals, commuters and explorers - that best describe the zonal visiting frequency and show that their composition varies considerably across users' place of residence and related demographics. We also identify 18 clusters of visiting spatial extent which form four groups that follow similar shapes of travel extent yet oriented in different directions. The approach proposed and applied in this study could be applied for any longitudinal individual travel demand data. 
\end{abstract}


\section{Introduction}

Human mobility has been known to exhibit some common features that extend beyond time and space. There is evidence to suggest that individual's daily travel time budget has been limited to approximately 70 minutes a day throughout human civilisation history and across cultures and geographies, whereas travel distances have increased as a result of technological advancements \citep{ahmed2014seventy}. Notwithstanding, the geographical area and the diversity of destinations visited, as well as the frequency of these visits vary considerably across the population. \cite{schneider2013unravelling} identify recurrent and distinctive motifs that represent individual daily mobility using travel survey and mobile phone data. Their analysis neglects the geographical aspects of individual mobility by limiting the representation of destinations to nodes in motif graphs. \cite{hasan2013spatiotemporal} develop a model for estimating macroscopic travel demand properties such as visiting frequency and origin-destination flows using smart card data. \cite{schlapfer2021universal} estimate the frequency of visits to different parts of the city as a function of travel distances and demonstrate that related variables follow universal laws. \cite{tu2018spatial} analyse spatial variations in passenger ridership in public transport and \cite{zhu2020understanding} compare the spatio-temporal heterogeneity in the usage of different shared mobility systems. There is thus a growing understanding of the macroscopic properties of human mobility based on disaggregate mobility traces. Notwithstanding, the same macroscopic properties may emerge from different compositions of mobility patterns at the individual level. 

The objective of this study is to unravel the commonalities and differences in the spatial properties of individual mobility patterns. In particular, our aim is to identify the extent to which users explore different parts of the network as well as identify distinctive user groups in terms of the spatial extent of their mobility patterns. We propose two representations of user profiles that are then subject to clustering. Study outputs can be used to devise targeted mobility products information and subscriptions. We apply the proposed analysis to a large-scale mobility data set from the public transport system in Stockholm, Sweden.

Multi-modal public transport networks are playing a critical role in urban mobility. Public transport systems are increasingly equipped with automated fare collection (AFC) systems which passively collect data concerning individual transactions. In the last decade, smart card data has been increasingly used to examine temporal daily and weekly variations \citep{ma2013mining, goulet2016inferring, ghaemi2017visual,deschaintres2019analyzing,he2020classification,egu2020investigating}. In contrast, little is known on market segments in regards to their geographical characteristics. The analysis of longitudinal travel data enables the investigation of how mobility patterns vary across the passengers' population and identify the spatial properties thereof. The approach proposed and applied in this study could be applied for any longitudinal individual travel demand data. 

In the following section we detail the input mobility data and the underlying network partitioning used in this study, followed by two approaches for user segmentation pertaining to zonal visiting profiles and the spatial extent of locations visited. In section 3 we provide a brief description of our application for the Stockholm County's public transport system. Next, we present and discuss the results of the proposed user segmentation clustering analysis for our application. We conclude with a discussion of study implications and suggestions for further research.

\section{Methodology}
In this section we describe the sequence of steps performed in order to unravel the spatial properties of individual mobility patterns. Two key ingredients are mobility data and a spatial partitioning of the geographical area under consideration. In the first two subsections we describe the features of the disaggregate longitudinal travel data required to enable the proposed analysis and the data-driven approach which we have adopted for generating travel demand zones. Next, we detail the two segmentation method proposed in this study. First a method based on a visiting profile which is agnostic to the specific locations visited is presented, followed by a method based on the spatial extent of the areas visited by users.

\subsection{Mobility input data}
\label{sec:Methodology_MobilityData}
Our analysis requires information on mobility records with longitudinal data on time and location stamps for the origin and destination of each journey for each traveller $i\in N$, Where $N$ is the set of all travellers considered in the analysis performed. Such data can be available for example from travel surveys, travel app tracking data, synthetic populations used in activity-based agent-based models, derived from car plate recognition records, ride-hailing or shared-fleet usage records. 

In the context of public transport journeys, observed origin and destination locations correspond to stops where the set of stops is denoted by $S$. A travel diary consisting of all journeys per traveller throughout the analysis period can be inferred from smart card data records by applying alighting location \citep{trepanier2007individual, munizaga2012estimation} and transfer inference algorithms \citep{gordon2013automated, yap2017robust}, depending on the fare validation scheme. Note that the analysis requires that the data contains card holder identifier which is consistent throughout the analysis period.  

For certain investigations it might be relevant to examine the socio- demographic characteristics of the obtained user classes. In some cases certain socio- demographic variables might be directly available from card holder information or based on the subscription/concession program (e.g. pensioners, students). However, some variables of interest may not be directly available at the individual passenger level. We therefore adopt the approach proposed by \cite{amaya2018estimating} and \cite{sari2019high} and adapted by \cite{SLProjectPart1} to identify the most likely home zone per traveller based on the frequency of each zone serving as an origin for each card holder. The zones used in this procedure can correspond to any zonal aggregation for which detailed socio-demographic (e.g. census) data is available for the relevant study area.


\subsection{Travel demand zones generation}
\label{sec:Methodology_ZonesGeneration}
Our analysis of users mobility patterns involves the identification of their visiting patterns of different parts of the geographical area under analysis. It is therefore essential to determine which set of zones will be used for describing passengers' visiting patterns. Zones can be based for example on those readily available from the official central bureau of statistics or by overlaying a grid and defining equal size grid cells. We hereby adopt a data-driven technique for generating travel demand zones that has been proposed and applied by \cite{LuoEtAl}.  

The aim of the adopted travel demand cluster generation technique is to cluster groups of stops based on information concerning both passenger flows and spatial distances. Hence, in addition to the passenger origin-destination matrix, the geographical coordinates of each stop is provided as input. The clustering of stops thus results with geographically compact zones (i.e. sets of stops) which exhibit similar travel demand patterns in terms of the distribution of travel destinations for journeys originating from stops included in the same zone. In the following we omit the time indexes because the analysis can be performed for any analysis period of choice without loss of generality.  

The method used to generate the demand zones follows the four-steps K-means-based station aggregation method proposed in \cite{LuoEtAl}. The four steps consist of:
\begin{enumerate}
    \item Use K-means to obtain clusters of stops. \\
    The clusters $\{C_k\}_{k=1,...,K}$ are a partition of $S$ with centers $\{\mu_k\}_{k=1,..,K}$. Consider geodesic distance $d(\cdot,\cdot)$ between stop $s\in S$ and cluster centers when performing the algorithm. Store the clustering results for a range of possible values of $k$. 
    \item Compute the distance-based metric. For each $k$ calculate: 
    \begin{itemize}
        \item Intra-cluster squared distance: $D^{intra}=\frac{1}{N} \sum\limits_{k=1}^{K}\sum\limits_{s\in C_k} d(\mu_k , s)^2$
        \item Inter-cluster squared distance: $D^{inter}=\min\limits_{k=1,..,K} \{d(\mu_k,s)^2\}$
    \end{itemize}
    With the aim of minimizing $D^{intra}$ and maximizing $D^{inter}$, we consider the problem of minimizing their ratio $\tau=\frac{D^{intra}}{D^{inter}}$
    \item Compute flow-based metric. For each {k=1,..,K} calculate:
    \begin{itemize}
        \item Intra-cluster flow: $F^{intra}=\frac{1}{k} \sum\limits_{k=1}^{k}\sum\limits_{s1,s2\in C_k} f(s1,s2)$
        \item Inter-cluster flow: $F^{inter}=\frac{1}{k\times(k-1)} \sum\limits_{k=1}^{K} \sum\limits_{l=1}^{K}
        \sum\limits_{\substack{s_i\in C_k1,s_j\in C_k2 \\ i\neq j}} f(s1,s2)$
        
        Where $f(s1,s2)$ is the observed passenger flow from stop $s1$ to stop $s2$ during the time period under consideration. 
    \end{itemize} 
    
    With the aim of maximizing intra-zonal flows and minimizing inter-zonal flows, we maximize the quantity $\delta=\frac{F^{intra}}{F^{inter}}$.
    \item Determine the number of clusters.\\
    Normalize the metrics to a $[0,1]$ range ($\tau \rightarrow \tau'$, $\delta \rightarrow \delta'$) and look for high values of the integrated metric $m(k)= \frac{\delta’}{\tau’}$ . $k^* = \arg\max\limits_K \{ m(k) \}$ gives the optimal number of clusters within our range.
\end{enumerate}

The resulting partitioning is such that all stops $s \in S$ are grouped into a $k$ number of zones denoted as $z \in Z$. Where the set of zones $Z$ is a collectively exhaustive and mutually exhaustive clustering of all the stops in $S$. Once the number of clusters is selected, we name and refer to each cluster of stops by the stop with the highest passenger volume included in each set.


\subsection{User segmentation based on zonal visiting profiles}
\label{sec:Methodology_ZonalVisitingProfiles}
Our goal is to identify groups of users that exhibit a similar behavior in terms of the frequency and diversity of zones visited. Note that in this user segmentation we are not interested in which specific zones are visited by individual users but rather the general properties of the zonal visiting profile. For example, if two individuals visit one zone 50\% of the time, i.e. have journeys for which this zone serves as the destination, 30\% for a second most-frequently visited zone and 20\% for their third most-frequently visited zone, then they are considered to have identical travel patterns for the sake of this analysis, regardless of the identity and locations of these zones. Hence, the key information needed for this analysis is to calculate for each individual user the number of journeys destined to each zone. The zones used here are those generated in the data-driven travel demand zone generation described in the previous section. 

For this analysis we consider the count of journeys ending in each zone for each user:
$W_{i,z}=\sum\limits_{s_i \in S}\sum\limits_{s_j \in z}{f_i(s_i,s_j)}$

Where $f_i(s_i,s_j)$ is the total flow, i.e. number of journeys, performed by user $i\in N$ with stop $s_i$ as an origin and stop $s_j$ as a destination. 

Each user $i\in N$ is characterized by a vector $W_i$ with each entry denoting the number of visits to each of the collectively exhaustive and mutually exclusive zones. These vectors are calculated and normalized for each user, so as to consider the share of the journeys destined to each zone rather than the absolute numbers.

We sort $W_i$ by descending values, i.e. from the most visited zone to the least visited. Consequently, we obtain an ordered and standardized vector per user which contains information about the visiting profile, where the first entry represents the percentage of journeys destined to the most visited zone, the second entry represents the percentage of journeys destined to the second most visited zone and so forth. This procedure results with \textit{zonal visiting frequency user profiles}.

Next, we cluster these profiles using the K-Means algorithm. The resulting centers from the K-Means algorithm are then interpreted based on the number of zones visited as well as the share of journeys attracted to each of the explored zones. According to the distribution of journeys share one can progressively define the clusters from the most to the least local.

The analysis can be further enriched by analyzing how the visiting profile clusters manifest themselves spatially, i.e. how does the share of users belonging to each visiting class varies geographically across home-zones as well as in relation to key socio-economic variables.

\subsection{User segmentation based on the spatial extent of locations visited}
\label{sec:Methodology_SpatialExtent}
While the analysis described in the previous section allows identifying clusters of users in terms of visiting patterns, it does not contain information on the spatial extent of user's mobility patterns and the shape thereof. In this subsequent step, we are interested in characterising users in terms of the spatial extent of the locations they visit during the course of the analysis period based on individual travel diary data. Unlike the visiting profile, the identity of the areas visited is of importance in this analysis. For each user's journey, we consider the origin and destination stops as two pinned locations. For this analysis, we choose to aggregate stops into grid cell zones by overlaying a grid over the study area, but this can be substituted by any other aggregation of choice. The grid cells avoids the problem of large variations in geographical size among zones. Each stop $s$ is linked to a grid cell/zone $z$. We then count for each zone how many journeys originated from this zone or destined to this zone. The user visit count per zone is thus
$
V_{i,s_x}=\sum\limits_{y \in S_x} f_i(s_x,s_y) + \sum\limits_{y \in S_x} f_i(s_y,s_x)
$

Each user visiting pattern is then represented using an array with each entree corresponding to the probability that a user visits this zone in relation to the overall visiting volume of the respective zone. To obtain this probability-representation, the data are normalized: $\tilde{V_i,s_x}=V_i,s_x/\sum\limits_{j=1}^{N}V_i,s_x$.
This procedure results with \textit{spatial visiting extent user profiles}. 

We then cluster users according to their normalised visit probability array. The clustering approach adopted is based on a Gaussian Mixture model, implemented using the Expectation Maximization algorithm. The choice of this method is mainly motivated by its capability of performing a soft classification, i.e. it considers the probability that a data point (user) belongs to each cluster and assigns it to the one for which it is most likely to belong (i.e. has the least distance from the centroid of the respective cluster).

\section{Application}
Stockholm County is comprised of 26 municipalities expanding over 6519.3 km$^2$ built over an archipelago and is home to 2.37 million inhabitants. Stockholm Region is the public transport authority overseeing all public transport services in the county. The multi-modal public transport system consists of bus, tram, metro, commuter train and ferry services. The network consists of more than 5,700 stops served by a rail network of 469 km and more than 9,000 km of bus service network. More than 1.2 million trips are performed on an average weekday by about 600,000 travellers. 

The fare scheme in Stockholm county involves tapping-in only. As part of a previous study, the virtual tap-out location was inferred for each trip and a transfer inference algorithm was applied. The details of these inference algorithms and a discussion of their validity and limitations are available at \cite{SLProjectPart1}. Consequently, given a disaggregate tap-in transaction database, a travel diary can be constructed for each card-holder where each entree is a journey performed during the course of the analysis period, containing the origin stop and (inferred) destination stop along with their respective time stamps. The output of this process constitutes the main input for our study. 

Our analysis is based on data from January 1, 2019 to December 31, 2019, with the exception of the month of July during which the data warehouse was under maintenance works. In total, 468,596,472 journeys were performed by 7,191,376 card holders during the 11-months analysis period. 

While no personal information is available at the card-holder level for this study, we use zonal-level socio-demographic and socio-economic data made available by the Swedish central bureau of statistics per census zone (there are 1,251 such zone in the case study area). As described in Section \ref{sec:Methodology_MobilityData}, we do so based on the inferred home zone per card holder. 

Given our interest in analyzing longitudinal data to study user behavioural patterns, we remove card-holders that use single tickets. This resulted in removing 38\% of the cards, associated with 20\% of all journeys. Consequently, we are left with a total of 4,423,783 remaining cards which performed 371,285,809 journey records. 

An additional filter is applied solely for the analysis described in Section \ref{sec:Methodology_ZonalVisitingProfiles}. Here users for which a home zone cannot be reasonably inferred are excluded. The total remaining users and journey records for this specific analysis are 3,782,954 and 368,710,217, respectively. Note that while this reflects a loss of 14\% of the users, only 0.68\% of the journeys are discarded, since frequent passengers are not affected by this filtering.

\section{Results and Analysis}
\subsection{Generating travel demand zones}
\begin{figure}[H]
    \centering
    \includegraphics[width=0.8\textwidth]{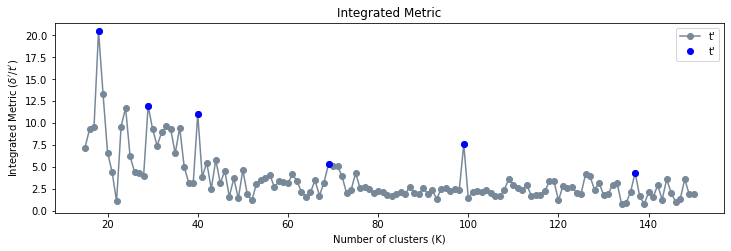}
    \caption{Integrated metric $m(k)$ value per number of clusters within the considered range.}
    \label{fig:KTHProj_01_2_IntegratedMetric}
\end{figure}

\begin{figure}[htp]
\sbox\twosubbox{%
  \resizebox{\dimexpr.99\textwidth-1em}{!}{%
    \includegraphics[height=3cm]{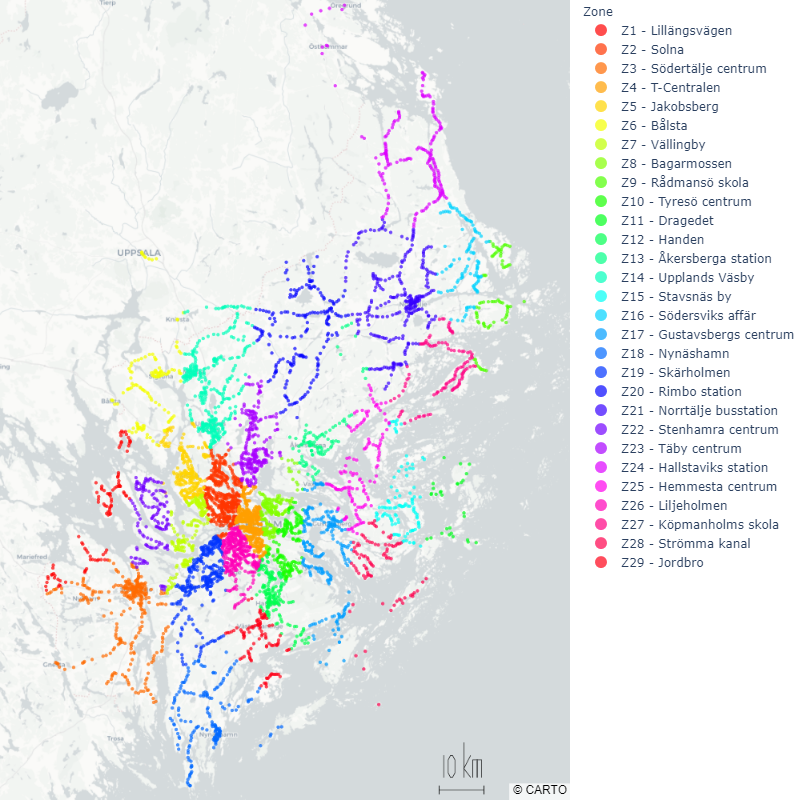}%
    \includegraphics[height=3cm]{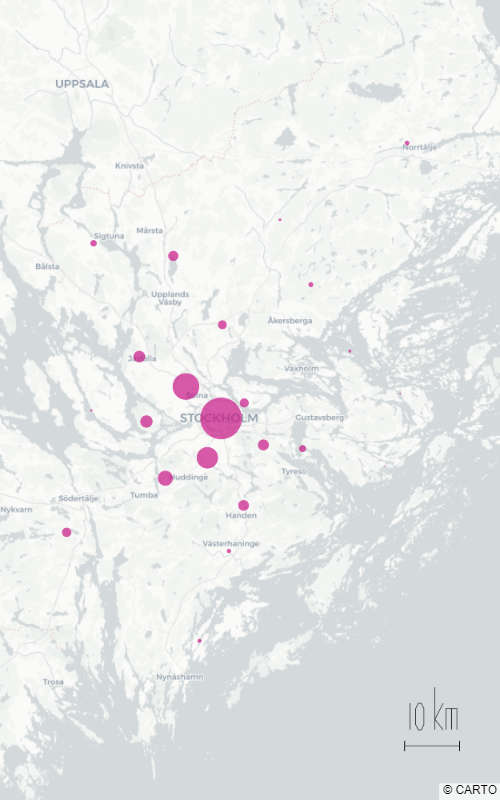}%
  }%
}
\setlength{\twosubht}{\ht\twosubbox}
\centering
\subcaptionbox{\label{f}}{%
  \includegraphics[height=\twosubht]{Figures/01_2/KTHProj_01_2_K29_MapStops.png}%
}\quad
\subcaptionbox{\label{s}}{%
  \includegraphics[height=\twosubht]{Figures/03_1/KTHPj_03_1_UsersPerHomeZone_Scatter_bigger.png}
}
\caption{Travel demand zones resulting from the 29 clusters (left), number of card holders residing in each zone based on our home-zone assignment results (right).}
\label{fig:KTHProj_01_2_K29_Map}
\end{figure}

The clustering procedure detailed in Section \ref{sec:Methodology_ZonesGeneration} has been applied for our case study. The values of the integrated metric, $m(k)$, for different number of clusters are shown in Figure \ref{fig:KTHProj_01_2_IntegratedMetric}. Even though 18 clusters yield the optimal metric value among all values tested (up to 150), we have chosen to opt for $k=29$ since it allows for a more nuanced analysis of geographical variations for our case study area and is the second-best local optima. In Figure \ref{fig:KTHProj_01_2_K29_Map} we display geographically the stop clustering results along with the respective number of card holders residing in each of the obtained zones based on the results of the home zone inference procedure described in Section \ref{sec:Methodology_MobilityData}.


\subsection{User exploration segments}
\begin{figure}[H]
     \centering
     \begin{subfigure}[b]{0.49\textwidth}
         \centering
         \includegraphics[width=\textwidth]{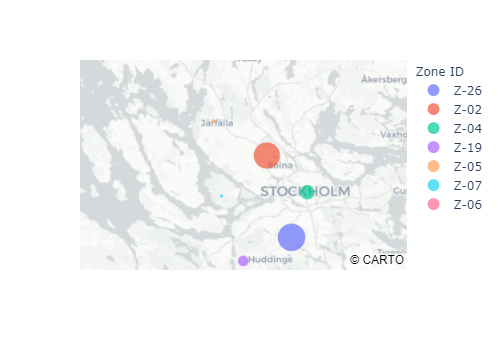}
     \end{subfigure}
     \hfill
     \begin{subfigure}[b]{0.49\textwidth}
         \centering
         \includegraphics[width=\textwidth]{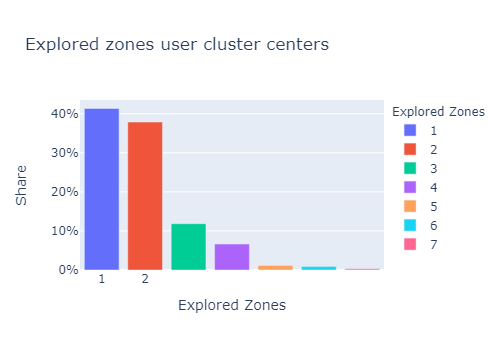}
     \end{subfigure}
        \caption{Geographical representation of a zonal visiting frequency user profile for a single traveller with larger nodes indicating zones visited more frequently (left) and the corresponding visiting frequency profile with zones presented in a descending order (right)}
        \label{fig:03_1_UserExample}
\end{figure}

We now turn to the clustering of users into distinctive exploration profiles. For each of the 3,782,954 cards included in our analysis we construct a vector consisting of 29 entrees, the number of zones for which stops have been clustered as described in the previous sub-section. Each of the entrees indicates the number of journeys for which the respective passenger has visited a given zone , i.e. this zone has served as a travel destination, during the course of the analysis period. For illustration purposes, we show in figure \ref{fig:03_1_UserExample} an example of a zonal visiting frequency user profiles for a selected traveller. 

\begin{figure}[H]
    \centering
    \includegraphics[width=0.7\textwidth]{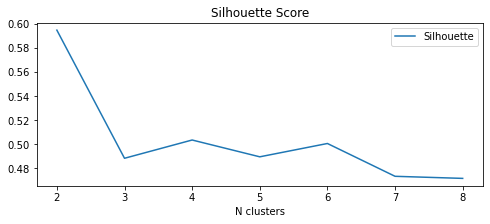}
    \caption{Silhouette index for number of clusters selection}
    \label{fig:KTHProj_03_1_Service_KM_NC_Selection}
\end{figure}

For each of the possible number of clusters considered, we have run the algorithm 30 times using different starting conditions in order to find the best performing partitioning. The average silhouette index values obtained for number of clusters of up to 8 are shown in Figure \ref{fig:KTHProj_03_1_Service_KM_NC_Selection}. We opt for three clusters in this case since, following the elbow rule-of-thumb, this constitutes a turning point with a local optimum. In addition, we found the results of three clusters to capture the most interesting patterns from an interpretative point of view.

\begin{figure}[H]
    \centering
    \includegraphics[width=0.8\textwidth]{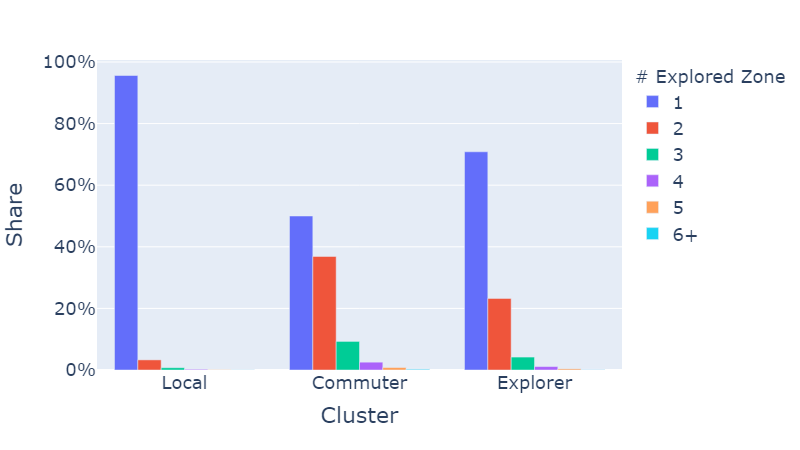}
    \caption{Zonal visiting profile per user exploration segment.}
    \label{fig:03_1_ZonalVisitingProfiles}
\end{figure}

The resulting cluster centers are presented in Figure \ref{fig:03_1_ZonalVisitingProfiles}. As described in Section \ref{sec:Methodology_ZonalVisitingProfiles}, the profiles are characterised in terms of the distribution of trips made over zones. We identify three user profiles with regard to their zonal visiting frequency profile:
\begin{itemize}
    \item \textbf{Local}: users which mostly (more than 90\% of the trips) travel to a single zone in the system. These users are mostly travelling within their home-zone area for a variety of trip purposes.
    \item \textbf{Commuter}: users that visit primarily two zones and for which the two most frequented zones are visited almost equally frequently (50\% and 40\% of the trips). These users are arguably likely to be commuters travelling back and forth between the home-zone and work-zone.
    \item \textbf{Explorer}: users for which there is a predominant destination zone (the destination of about 70\% of their trips), but they still visit several other regions of the city to various extents. 
\end{itemize}
The plurality of the card holders belong to the 'Commuter' cluster with 39.1\% of all of the card holders, followed by 'Local' with 32.3\% and only 28.6\% of the card holders categorised as an 'Explorer'.

\begin{figure}[H]
    \centering
    \includegraphics[width=\textwidth]{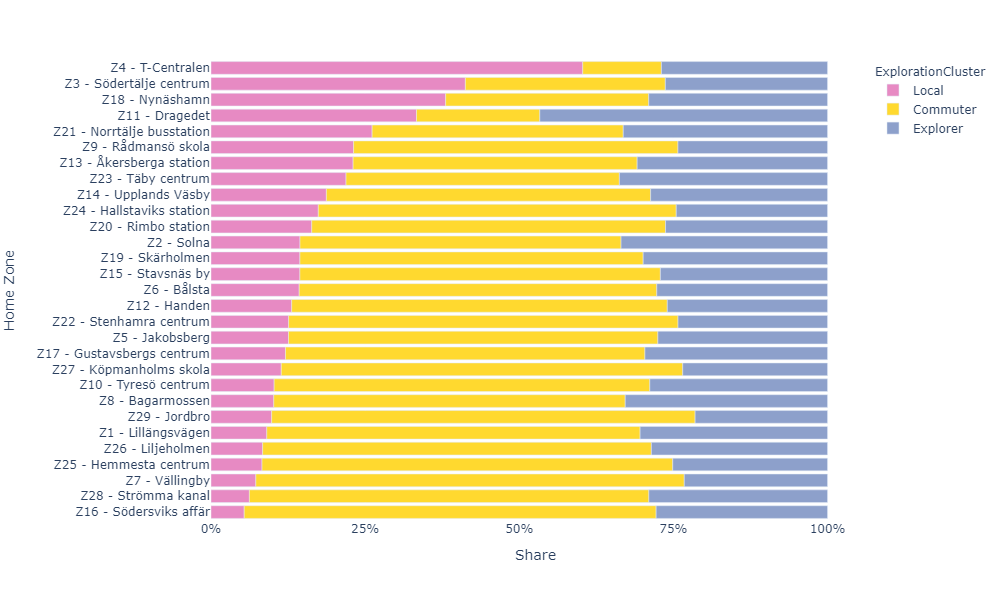}
    \caption{Share of user visiting pattern segment per home-zone area}
    \label{fig:KTHPj_03_1_ZonalVisitingProfilePerHomeZone}
\end{figure}

We then turn to analysing the composition of user segments per home-zone. The share of each segment amongst card-holders residing in each of the zones is shown in Figure \ref{fig:KTHPj_03_1_ZonalVisitingProfilePerHomeZone}. The reader may refer to Figure \ref{fig:KTHProj_01_2_K29_Map} in order to geographically identify the zone using the displayed zone number and name. As can be expected, areas with high job intensity such as the central parts of Stockholm city (Z-4 T-Centralen, service and retail), Södertälje (z-3, manufacturing) and Nynäshamn (z-18, harbor), have a higher share of 'Local' users. The 'Commuter' type is predominant for those areas that are close to an urban center, e.g. Z-7, Vällingby and Z-29, Jordbro. The share of users belonging to the 'Explorer' type is overall evenly distributed across zones, with a peak for users residing in the archipelago areas (i.e. z-11, Dragedet).

Next, we examine key socio-demographic attributes of users assigned to each of the three visiting profile clusters. Note that this information is obtained from the underlying census zones. Summary statistics are summarised in Table \ref{tab:KTHPj_03_1_SocioEconomic}. As can be seen, users following a 'Commuter' visiting pattern reside in areas characterised by a a lower social index which reflects an array of social variables, lower than average income level and a higher than average share of residents from foreign background. The opposite holds for 'Local' users. 'Explorer' users reside in areas which are on par with the overall mean values across the case study area.

\begin{table}[htbp]
  \centering
    \begin{tabular}{|l|r|r|r|r|}
    \hline
    \multicolumn{1}{|c|}{\textbf{Attribute}} & \multicolumn{1}{c|}{\textbf{Commuter}} & \multicolumn{1}{c|}{\textbf{Explorer}} & \multicolumn{1}{c|}{\textbf{Local}} & \multicolumn{1}{c|}{\textbf{Mean}} \\
    \hline
    Income (annual in '000 SEK) & 269 & 280 & 299 & 281 \\
    Social Index (1-10) & 7.74 & 8.44 & 9.09 & 8.35 \\
    Foreign-background & 12.56\% & 11.86\% & 9.87\% & 11.58\% \\
    \hline
    \end{tabular}%
  \caption{Socio-economic attributes per user visiting profile}
  \label{tab:KTHPj_03_1_SocioEconomic}%
\end{table}%

\subsection{User segments by travel pattern spatial extent}
\begin{figure}[H]
    \centering 
  \includegraphics[width=0.5\linewidth]{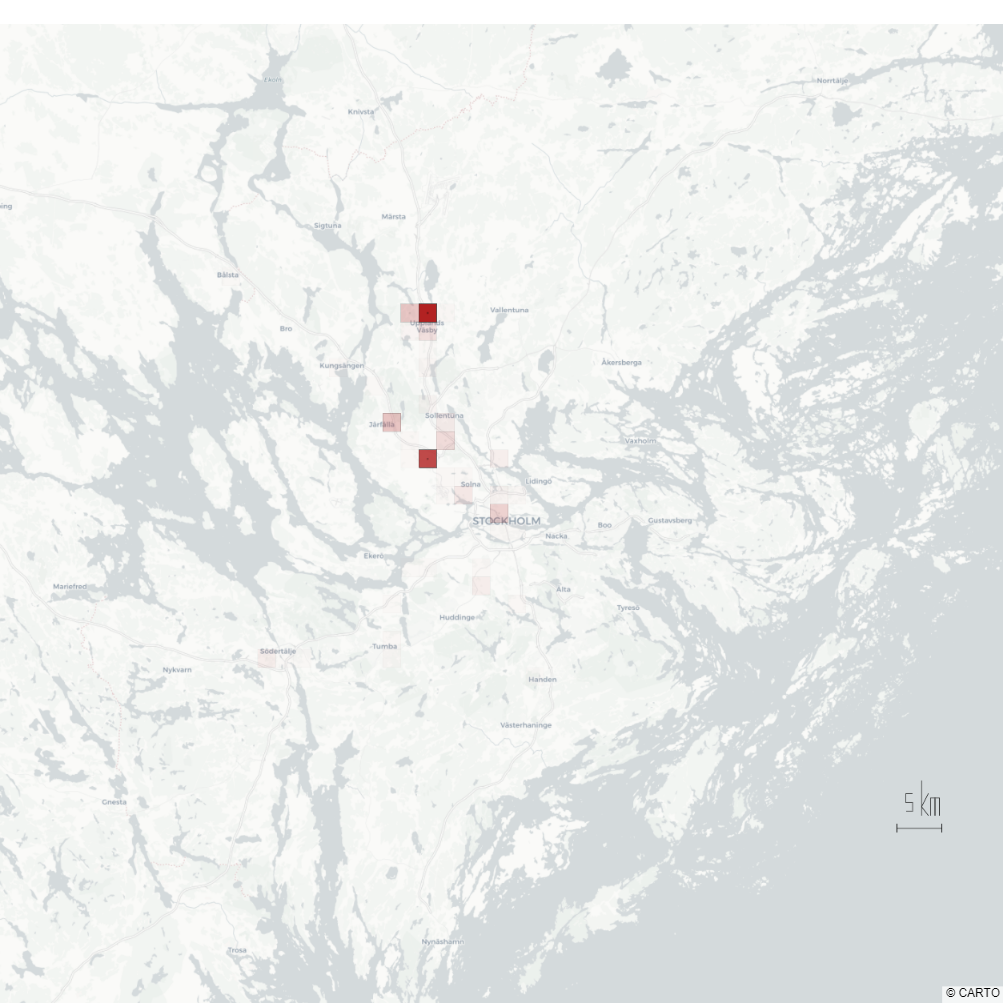}
\caption{Visiting heat map representation of for an individual traveller over the 2km by 2km pixels grid.}
\label{fig:KTHPj_03_4_2km_user_ck375109_DensityMap}
\end{figure}

We follow the method described in Section \ref{sec:Methodology_SpatialExtent} in order to analyze the spatial extent of the area visited by each user. For each of the 4,423,783 cards included in this analysis we generate a spatial visiting extent user profile with the share of journeys destined to each spatial unit considered. We illustrate this using a heat map representing the share of visits per grid cell as shown in \ref{fig:KTHPj_03_4_2km_user_ck375109_DensityMap} for an example user. 

For the analysis of spatial analysis extent, we choose to overlay a grid over the case study area. On one hand, a more finely meshed grid implies a larger number of cells, posing the risk of over-fitting and scarcity of stops per cell in peripheral parts of the network. On the other hand, a more dispersed grid would instead bundles many stops within the same cell, and may compromise geographical nuances. We have experimented with various grid sizes ranging from 0.5km on 0.5km to 2.5km on 2.5km with 0.5km increments. in addition, the type of covariance matrix to adopt for the Gaussian Mixture Model has to be specified.

To evaluate the different configurations we calculate for each of them the BIC and the AIC of 10 randomly sampled 10\% subsets of the data, based on which we have selected using grid cells of 2km on 2km. The configuration using a mesh with $2 \times 2$ km cells and a diagonal type of co-variance yielded a total of 15 clusters. 

\begin{figure}[H]
    \centering 
\begin{subfigure}{0.4\textwidth}
    \includegraphics[width=\textwidth]{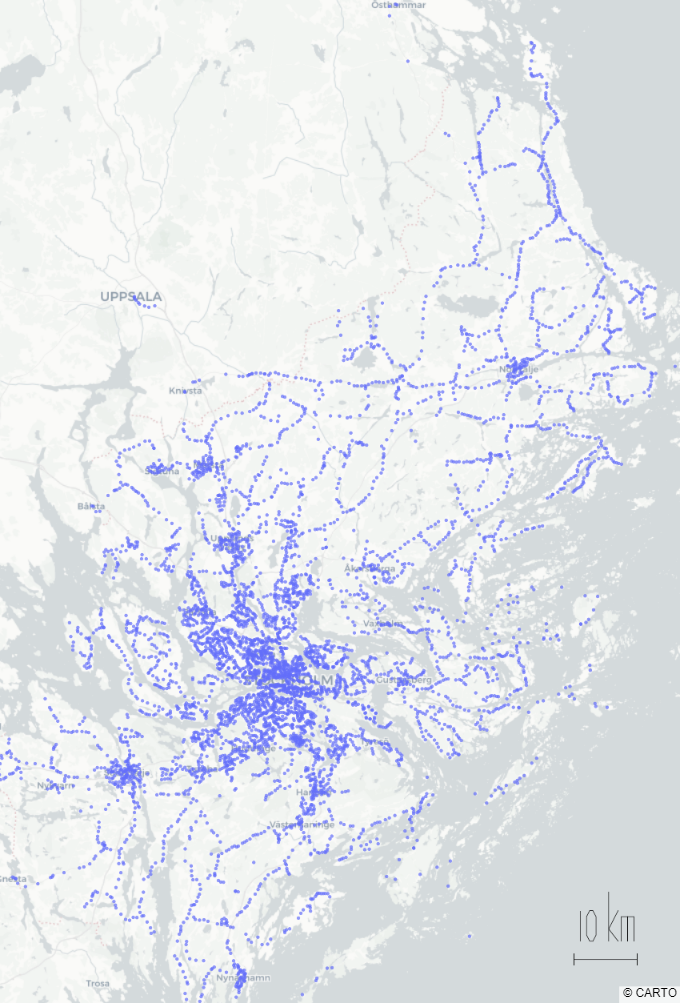}
    \caption{}
    \label{fig:KTHPj_04_3_MapEmpty}
\end{subfigure}\hfil 
\begin{subfigure}{0.4\textwidth}
  \includegraphics[width=\linewidth]{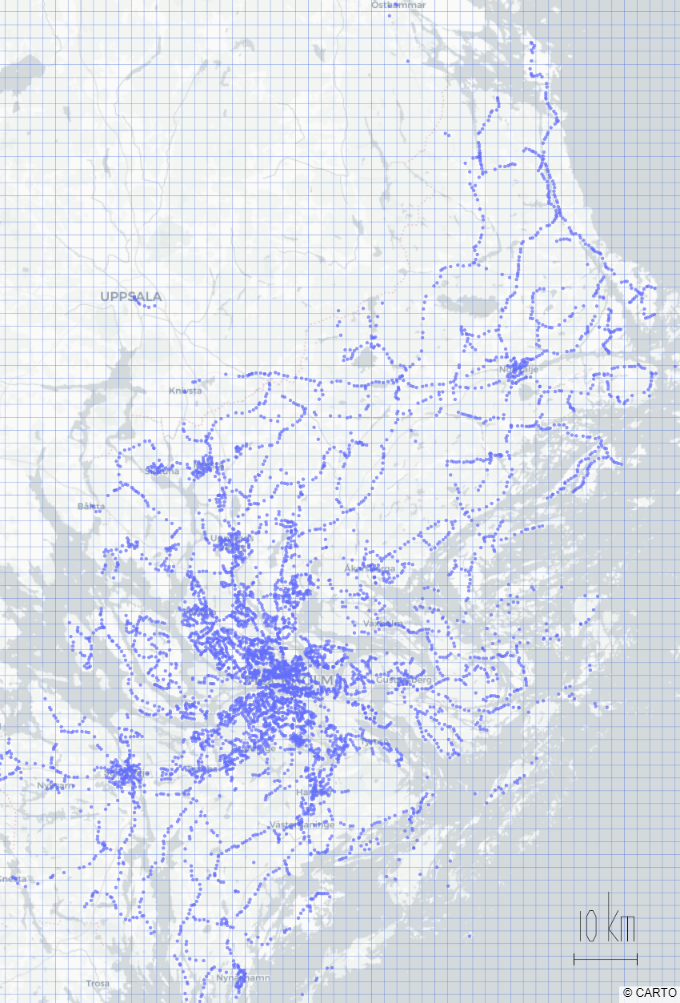}
    \caption{}
    \label{fig:KTHPj_04_3_MapPixelsGrid}
\end{subfigure}
\caption{Overlaying a 2km on 2 km grid on the Stockholm County.}
\label{fig:KTHPj_04_3_Map}
\end{figure}

With this size, the grid covering the entire case study consists of $54 \times 86$ cells (see \ref{fig:KTHPj_04_3_Map}. The number of active cells in the grid, i.e. the number of cells where at least one stop of the network is situated, is 1,154, whereas the total number of cells is 4,644 (largely due to the large bodies of water in our case study area). 

\begin{figure}[H]
\centering 
\begin{subfigure}{0.33\textwidth}
  \includegraphics[width=\linewidth]{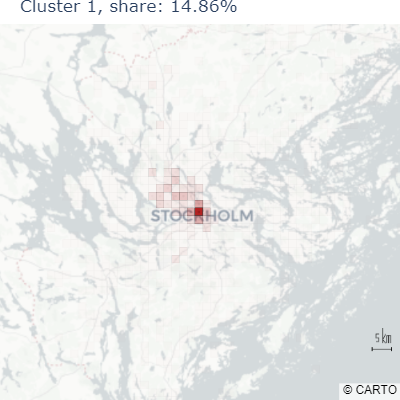}
\end{subfigure}\hfil 
\begin{subfigure}{0.33\textwidth}
  \includegraphics[width=\linewidth]{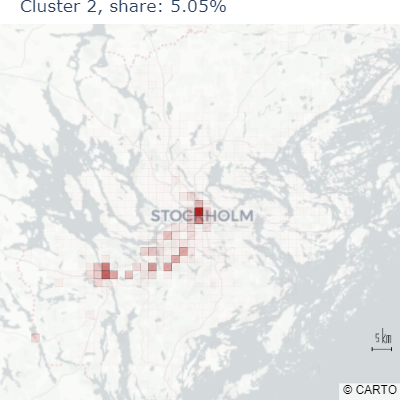}
\end{subfigure}
\begin{subfigure}{0.33\textwidth}
  \includegraphics[width=\linewidth]{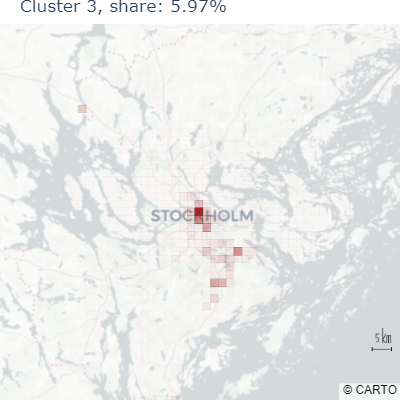}
\end{subfigure}

\medskip
\begin{subfigure}{0.33\textwidth}
  \includegraphics[width=\linewidth]{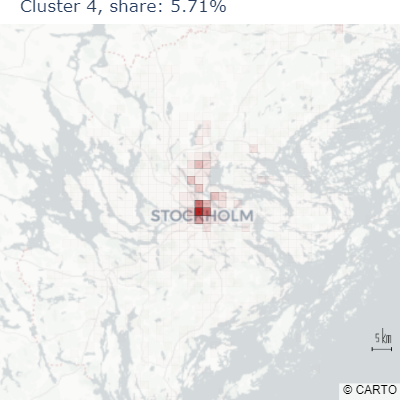}
\end{subfigure}\hfil 
\begin{subfigure}{0.33\textwidth}
  \includegraphics[width=\linewidth]{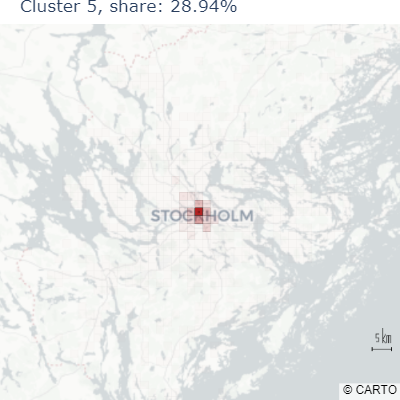}
\end{subfigure}
\begin{subfigure}{0.33\textwidth}
  \includegraphics[width=\linewidth]{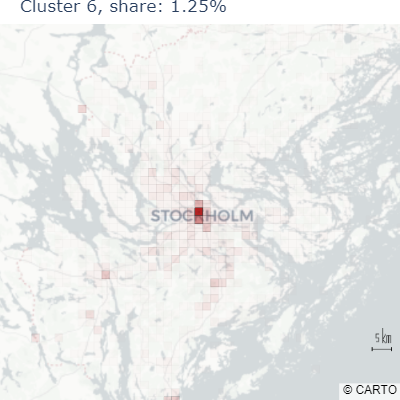}
\end{subfigure}

\medskip
\begin{subfigure}{0.33\textwidth}
  \includegraphics[width=\linewidth]{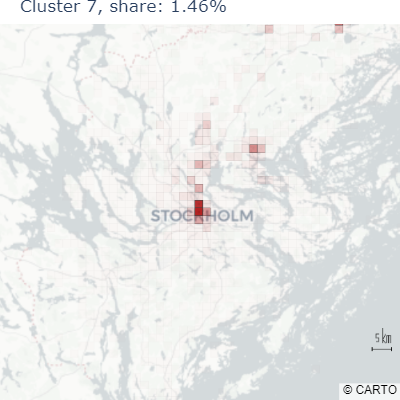}
\end{subfigure}\hfil 
\begin{subfigure}{0.33\textwidth}
  \includegraphics[width=\linewidth]{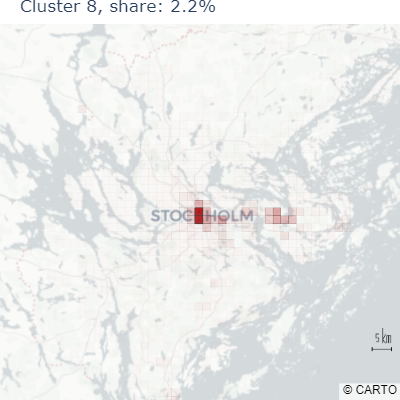}
\end{subfigure}
\begin{subfigure}{0.33\textwidth}
  \includegraphics[width=\linewidth]{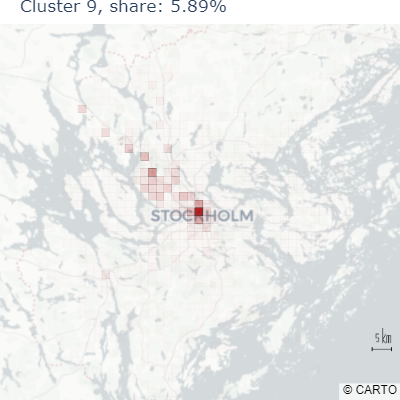}
\end{subfigure}

\medskip
\begin{subfigure}{0.33\textwidth}
  \includegraphics[width=\linewidth]{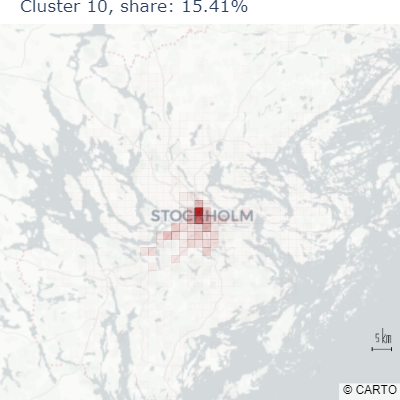}
\end{subfigure}\hfil 
\begin{subfigure}{0.33\textwidth}
  \includegraphics[width=\linewidth]{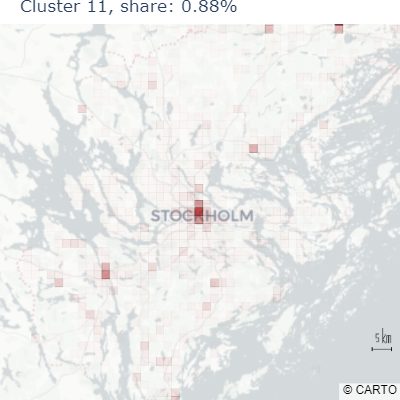}
\end{subfigure}
\begin{subfigure}{0.33\textwidth}
  \includegraphics[width=\linewidth]{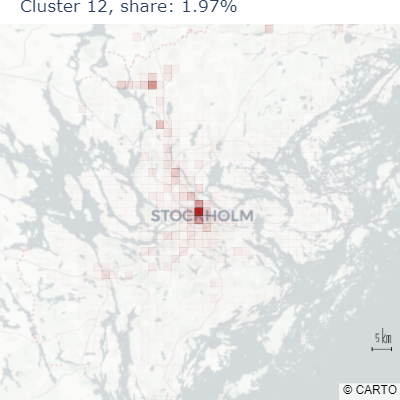}
\end{subfigure}

\medskip
\begin{subfigure}{0.33\textwidth}
  \includegraphics[width=\linewidth]{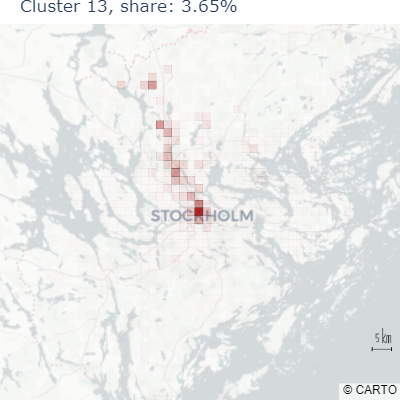}
\end{subfigure}\hfil 
\begin{subfigure}{0.33\textwidth}
  \includegraphics[width=\linewidth]{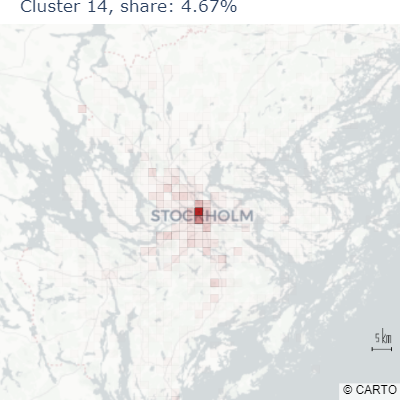}
\end{subfigure}
\begin{subfigure}{0.33\textwidth}
  \includegraphics[width=\linewidth]{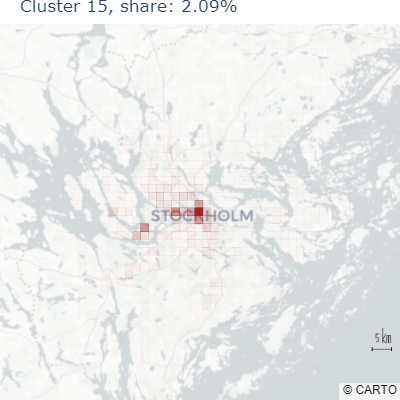}
\end{subfigure}
\caption{Visiting heat maps of the 18 identified user segments}
\label{fig:KTHPj_03_4_K15Clusters_DensityMap}
\end{figure}

The mean of the Gaussians determining the centers of the 18 clusters are shown in Figure \ref{fig:KTHPj_03_4_K15Clusters_DensityMap}, along with the share of users assigned to each cluster. The city center of Stockholm is clearly dominant in all clusters, in line with the highly monocentric structure characterising the structure of Stockholm metropolitan area despite the recent emergence of secondary activity centers \citep{cats2015identification}. 

In the following we discuss key observations from the clusters obtained. The largest cluster, gathering almost 29\% of the users (more than one out of four cards), is cluster 5. This cluster represents users that the spatial extent of their travel patterns is almost entirely limited to central areas of Stockholm city - 66\% of their journeys are confined to 7 neighbouring zones, i.e. 28 km$^2$.

We observe that the remaining clusters can be loosely described as following into one of three types. Several clusters exhibit similar shapes of travel extent yet oriented in different directions. For example, cluster 2 corresponds to users that travel between central parts of Stockholm and the city of Södertalje (south-west end of Stockholm County) as well as visit areas situated along the corridors connecting these cities. A similar pattern emerges for cluster 8, but instead of Södertälje the spatial extent profile is oriented towards Gustavsberg-Värmdö. Other clusters following this pattern include cluster 7 (Stockholm - Norrtälje direction), cluster 9 (Stockholm - Jakobsberg-Bålsta), and cluster 13 (Stockholm-Märsta).

Another group of clusters includes those exhibiting more sparse visiting patterns, forming a cloud shape with a variety of locations visited more uniformly, yet mostly confined to a limited part of the case study area. The following clusters fall in this category: cluster 1 (North-West area of Stockholm city - Solna), cluster 3 (South-East area of Stockholm city - Tyresö, Handen), cluster 4 (North of Stockholm - Danderyd, Täby), cluster 10 (South-West area of Stockholm - Skarholmen, Huddinge, Farsta), and cluster 15 (West - Ekerö, Bromma)

The last group of clusters includes those that exhibit a more disperse pattern, characterized by scattered visits to non-neighboring parts of the case study area: cluster 6, cluster 11, cluster 12, cluster 14.

\section{Conclusion}
We proposed two methods for clustering travellers based on the spatial properties of their mobility patterns using longitudinal travel data. These methods were applied for the case of smart card data from the multi-modal public transport system of Stockholm County. After partitioning the network based on a data-driven approach, we represent users patterns in terms of zonal visiting frequency profiles and grid-cells spatial extent heatmaps. We identify three clusters - denominated locals, commuters and explorers - that best describe the zonal visiting frequency and show that their composition varies considerably across users' place of residence and related demographics. We also unravel 18 clusters of visiting spatial extent which form four groups that follow the same overall trend in terms of intensity and concentration of the zones visited, yet prevalent in different locations across the network. 

The user segmentation approach proposed in this study can be used to devise products and fare schemes that cater for specific mobility patterns in terms of frequency and extent of locations visited across the network. Moreover, the analysis can be used to identify gaps in the network that may benefit from increased accessibility. Future research may investigate how the identified segments have evolved over a long period of times, for example in relation to policies such as stimulating a more polycentric regional development. This can be especially insightful in understanding changes in mobility patterns and user segmentation as a consequence of major network developments or significant disruptions such as the COVID-19 pandemic.

\section*{Acknowledgements}
This study is funded by Region Stockholm, project "Unravelling travel demand patterns using Access card data" RS 2019-0499. We also thank Region Stockholm for providing the smart card data that made this study possible. The authors also thank Isak Rubensson, Matej Cebecauer and Erik Jenelius for their support in the process.

\bibliographystyle{apa-good}
\bibliography{references}

\begin{thebibliography}{21}
\expandafter\ifx\csname natexlab\endcsname\relax\def\natexlab#1{#1}\fi
\expandafter\ifx\csname url\endcsname\relax
  \def\url#1{{\tt #1}}\fi
\expandafter\ifx\csname urlprefix\endcsname\relax\def\urlprefix{URL }\fi

\bibitem[{Ahmed \& Stopher(2014)}]{ahmed2014seventy}
Ahmed, A., \& Stopher, P. (2014).
\newblock Seventy minutes plus or minus 10—a review of travel time budget
  studies.
\newblock {\em Transport Reviews\/}, {\em 34\/}(5), 607--625.

\bibitem[{Amaya et~al.(2018)Amaya, Cruzat, \& Munizaga}]{amaya2018estimating}
Amaya, M., Cruzat, R., \& Munizaga, M.~A. (2018).
\newblock Estimating the residence zone of frequent public transport users to
  make travel pattern and time use analysis.
\newblock {\em Journal of Transport Geography\/}, {\em 66\/}, 330--339.

\bibitem[{Cats et~al.(2015)Cats, Wang, \& Zhao}]{cats2015identification}
Cats, O., Wang, Q., \& Zhao, Y. (2015).
\newblock Identification and classification of public transport activity
  centres in stockholm using passenger flows data.
\newblock {\em Journal of Transport Geography\/}, {\em 48\/}, 10--22.

\bibitem[{Deschaintres et~al.(2019)Deschaintres, Morency, \&
  Tr{\'e}panier}]{deschaintres2019analyzing}
Deschaintres, E., Morency, C., \& Tr{\'e}panier, M. (2019).
\newblock Analyzing transit user behavior with 51 weeks of smart card data.
\newblock {\em Transportation Research Record\/}, {\em 2673\/}(6), 33--45.

\bibitem[{Egu \& Bonnel(2020)}]{egu2020investigating}
Egu, O., \& Bonnel, P. (2020).
\newblock Investigating day-to-day variability of transit usage on a multimonth
  scale with smart card data. a case study in lyon.
\newblock {\em Travel Behaviour and Society\/}, {\em 19\/}, 112--123.

\bibitem[{Ghaemi et~al.(2017)Ghaemi, Agard, Tr{\'e}panier, \&
  Partovi~Nia}]{ghaemi2017visual}
Ghaemi, M.~S., Agard, B., Tr{\'e}panier, M., \& Partovi~Nia, V. (2017).
\newblock A visual segmentation method for temporal smart card data.
\newblock {\em Transportmetrica A: Transport Science\/}, {\em 13\/}(5),
  381--404.

\bibitem[{Gordon et~al.(2013)Gordon, Koutsopoulos, Wilson, \&
  Attanucci}]{gordon2013automated}
Gordon, J.~B., Koutsopoulos, H.~N., Wilson, N. H.~M., \& Attanucci, J.~P.
  (2013).
\newblock Automated inference of linked transit journeys in london using
  fare-transaction and vehicle location data.
\newblock {\em Transportation research record\/}, {\em 2343\/}(1), 17--24.

\bibitem[{Goulet-Langlois et~al.(2016)Goulet-Langlois, Koutsopoulos, \&
  Zhao}]{goulet2016inferring}
Goulet-Langlois, G., Koutsopoulos, H.~N., \& Zhao, J. (2016).
\newblock Inferring patterns in the multi-week activity sequences of public
  transport users.
\newblock {\em Transportation Research Part C: Emerging Technologies\/}, {\em
  64\/}, 1--16.

\bibitem[{Hasan et~al.(2013)Hasan, Schneider, Ukkusuri, \&
  Gonz{\'a}lez}]{hasan2013spatiotemporal}
Hasan, S., Schneider, C.~M., Ukkusuri, S.~V., \& Gonz{\'a}lez, M.~C. (2013).
\newblock Spatiotemporal patterns of urban human mobility.
\newblock {\em Journal of Statistical Physics\/}, {\em 151\/}(1), 304--318.

\bibitem[{He et~al.(2020)He, Agard, \& Tr{\'e}panier}]{he2020classification}
He, L., Agard, B., \& Tr{\'e}panier, M. (2020).
\newblock A classification of public transit users with smart card data based
  on time series distance metrics and a hierarchical clustering method.
\newblock {\em Transportmetrica A: Transport Science\/}, {\em 16\/}(1), 56--75.

\bibitem[{Kholodov et~al.(2021)Kholodov, Jenelius, Cats, {van Oort}, Mouter,
  Cebecauer, \& Vermeulen}]{SLProjectPart1}
Kholodov, Y., Jenelius, E., Cats, O., {van Oort}, N., Mouter, N., Cebecauer,
  M., \& Vermeulen, A. (2021).
\newblock Public transport fare elasticities from smartcard data: Evidence from
  a natural experiment.
\newblock {\em Transport Policy\/}, {\em 105\/}, 35--43.

\bibitem[{Luo et~al.(2017)Luo, Cats, \& Lint}]{LuoEtAl}
Luo, D., Cats, O., \& Lint, J. (2017).
\newblock Constructing transit origin–destination matrices with spatial
  clustering.

\bibitem[{Ma et~al.(2013)Ma, Wu, Wang, Chen, \& Liu}]{ma2013mining}
Ma, X., Wu, Y., Wang, Y., Chen, F., \& Liu, J. (2013).
\newblock Mining smart card data for transit riders’ travel patterns.
\newblock {\em Transportation Research Part C: Emerging Technologies\/}, {\em
  36\/}, 1--12.

\bibitem[{Munizaga \& Palma(2012)}]{munizaga2012estimation}
Munizaga, M.~A., \& Palma, C. (2012).
\newblock Estimation of a disaggregate multimodal public transport
  origin--destination matrix from passive smartcard data from santiago, chile.
\newblock {\em Transportation Research Part C: Emerging Technologies\/}, {\em
  24\/}, 9--18.

\bibitem[{Sari~Aslam et~al.(2019)Sari~Aslam, Cheng, \& Cheshire}]{sari2019high}
Sari~Aslam, N., Cheng, T., \& Cheshire, J. (2019).
\newblock A high-precision heuristic model to detect home and work locations
  from smart card data.
\newblock {\em Geo-spatial Information Science\/}, {\em 22\/}(1), 1--11.

\bibitem[{Schl{\"a}pfer et~al.(2021)Schl{\"a}pfer, Dong, O’Keeffe, Santi,
  Szell, Salat, Anklesaria, Vazifeh, Ratti, \& West}]{schlapfer2021universal}
Schl{\"a}pfer, M., Dong, L., O’Keeffe, K., Santi, P., Szell, M., Salat, H.,
  Anklesaria, S., Vazifeh, M., Ratti, C., \& West, G.~B. (2021).
\newblock The universal visitation law of human mobility.
\newblock {\em Nature\/}, {\em 593\/}(7860), 522--527.

\bibitem[{Schneider et~al.(2013)Schneider, Belik, Couronn{\'e}, Smoreda, \&
  Gonz{\'a}lez}]{schneider2013unravelling}
Schneider, C.~M., Belik, V., Couronn{\'e}, T., Smoreda, Z., \& Gonz{\'a}lez,
  M.~C. (2013).
\newblock Unravelling daily human mobility motifs.
\newblock {\em Journal of The Royal Society Interface\/}, {\em 10\/}(84),
  20130246.

\bibitem[{Tr{\'e}panier et~al.(2007)Tr{\'e}panier, Tranchant, \&
  Chapleau}]{trepanier2007individual}
Tr{\'e}panier, M., Tranchant, N., \& Chapleau, R. (2007).
\newblock Individual trip destination estimation in a transit smart card
  automated fare collection system.
\newblock {\em Journal of Intelligent Transportation Systems\/}, {\em 11\/}(1),
  1--14.

\bibitem[{Tu et~al.(2018)Tu, Cao, Yue, Zhou, Li, \& Li}]{tu2018spatial}
Tu, W., Cao, R., Yue, Y., Zhou, B., Li, Q., \& Li, Q. (2018).
\newblock Spatial variations in urban public ridership derived from gps
  trajectories and smart card data.
\newblock {\em Journal of Transport Geography\/}, {\em 69\/}, 45--57.

\bibitem[{Yap et~al.(2017)Yap, Cats, van Oort, \& Hoogendoorn}]{yap2017robust}
Yap, M., Cats, O., van Oort, N., \& Hoogendoorn, S. (2017).
\newblock A robust transfer inference algorithm for public transport journeys
  during disruptions.
\newblock {\em Transportation research procedia\/}, {\em 27\/}, 1042--1049.

\bibitem[{Zhu et~al.(2020)Zhu, Zhang, Kondor, Santi, \&
  Ratti}]{zhu2020understanding}
Zhu, R., Zhang, X., Kondor, D., Santi, P., \& Ratti, C. (2020).
\newblock Understanding spatio-temporal heterogeneity of bike-sharing and
  scooter-sharing mobility.
\newblock {\em Computers, Environment and Urban Systems\/}, {\em 81\/}, 101483.

\end{thebibliography}
\end{document}